\documentclass[psfig]{article}
\usepackage{graphicx}
\usepackage{dcolumn}
\usepackage{bm}
\begin{document}
\setcounter{page}{1}
\begin{center}
{\Large \bf Possible $\eta$-Mesic $^3$He States within the 
Finite Rank Approximation}
\vskip0.5cm
{N. G.~Kelkar$^1$, K. P. Khemchandani$^{2,3}$ and B. K. Jain$^3$}
\vskip0.5cm
{\it 
{$^1$Departamento de Fisica, Universidad de los Andes, \\
Cra 1E, 18A-10, Bogota, A. A. 4976, Colombia\\
$^2$ Dept. de Fisica Teorica and IFIC, 
Centro Mixto Univ. de Valencia-CSIC\\
Institutos de Investigacion de Paterna, Aptd. 22085, 46071 Valencia, Spain\\
$^3$Department of Physics, University of Mumbai, 
Mumbai, India}}
\end{center}

\begin{abstract}
We extend the method of time delay proposed by Eisenbud and Wigner, 
to search for unstable states formed by $\eta$ mesons and the $^3$He nucleus.
Using few body equations to describe $\eta$-$^3$He elastic scattering, 
we predict resonances and unstable bound states within different
models of the $\eta N$ interaction. The $\eta\,^3$He states predicted 
within this novel approach are in agreement with the recent claim of 
the evidence of $\eta$-mesic $^3$He made by the TAPS collaboration. 
\end{abstract}
\vskip0.5cm
\noindent
PACs numbers:{14.40.Aq, 03.65.Nk, 14.20.Gk}
\vskip0.5cm
A claim of positive evidence for the existence of the long sought after
$\eta$-mesic nuclei was made last year by the TAPS collaboration \cite{taps} 
on the basis of their investigations on the photoproduction reaction, 
$\gamma \,^3$He $\rightarrow \eta X$.   
This indeed comes as a big step in the understanding of many 
nuclear and particle physics related problems of fundamental interest.  
We list here the connection of $\eta$-physics to a few of these
important issues: (i) the $\eta$-N interaction is 
dominated by the coupling to the $N^*(1535)$ resonance and hence, studies of 
the $\eta$ meson in the nuclear medium could yield information on the
$\eta N N^*$ coupling constant. (ii) The isospin and charge
symmetry breaking (CSB) in quantum chromodynamics (QCD) arises due to the
difference in the up and down quark masses due to their electromagnetic
interaction. The observation of the $\pi$-$\eta$-$\eta^{\prime}$ mixing
provides one of the best possibilities to study directly the CSB effects 
\cite{machner}. 
(iii) The relative cross sections for the reactions $\eta p$, 
$\eta^{\prime} p \, \rightarrow \eta p$, $\eta^{\prime} p$ and 
$\pi^- p \rightarrow \eta n$, $\eta^{\prime} n$, provide a sensitive 
test for the presence of a strange-antistrange ($s \bar{s}$) component 
in the nucleon's wave function \cite{dover}.  
Obviously then, this evidence of an $\eta$-mesic 
$^3$He increased the interest in this field which started about a decade 
ago \cite{bhal,liu}. We therefore found it timely to 
search for the $\eta$-mesic resonances using the method of
time delay proposed by Eisenbud and Wigner \cite{eiswig} 
about half a century ago, but revived only recently to 
successfully characterize the meson and baryon resonances 
\cite{weall,weall2}. 
From our experience in \cite{weall}, 
we found that with this method (which is mentioned in text books
and literature as a necessary condition for the existence of a
resonance) 
we not only confirmed the existing resonances, but also found new ones.
Our finding of the exotic pentaquark resonance 
around 1540 MeV from $K^+N$ data \cite{weall2} was 
indeed confirmed by several recent experiments (listed in \cite{weall2}).

The time delay in 
$\eta \,^3$He elastic scattering is evaluated using few-body equations for
the $\eta \,^3$He system involving inputs from different models of 
the elementary $\eta N$ interaction. In an earlier work \cite{etawe}, 
the near threshold data on the $p d \rightarrow \,^3$He $\eta$ reaction
were well reproduced using the same few body equations to describe 
the $\eta \,^3$He final state interaction.      
In what follows, the method of time delay is briefly introduced, 
followed by an example which demonstrates the usefulness of this 
concept even at negative energies.

Back in the early fifties, Eisenbud and Wigner 
related the energy derivative of
the phase shift to the time delay in scattering as, 
$\Delta t(E) = 2 \,\hbar d \delta / dE$. 
A more useful definition, namely, the time delay matrix
was given by Eisenbud and elaborated by Smith \cite{smith}. 
An element of this matrix, $\Delta t_{ij}$, which is the
time delay in the emergence of a particle in the $j^{th}$ channel after
being injected in the $i^{th}$ channel is given by,
\begin{equation}\label{2}
\Delta t_{ij} = \Re e \biggl [ -i \hbar (S_{ij})^{-1} {dS_{ij} \over dE}
\biggr ] \, ,
\end{equation}
where $S_{ij}$ is an element of the corresponding $S$-matrix. In an
eigenphase formulation of the $S$-matrix, one can easily see that the time
delay as defined above in (\ref{2}) is the energy derivative of the 
phase shift. 
Alternatively, writing the $S$-matrix in terms of the $T$-matrix as,
${\bf S} = 1 + 2\,i\,{\bf T}$,
one can evaluate time delay in terms of the $T$-matrix.
The time delay in elastic scattering, i.e. $\Delta t_{ii}$, is given in terms
of the $T$-matrix as,
\begin{equation}\label{7}
S^*_{ii} \,S_{ii}\, \Delta t_{ii}(E)\, =\, 2 \,\hbar\,
\biggl[ \Re e \biggl({dT_{ii} \over dE}\biggr)\,+ \,2 \,\Re e T_{ii}\,\,
\Im m \biggl ({dT_{ii} \over dE}\biggr) \,-\, 2\, \Im m T_{ii}\,\,
\Re e\biggl( {dT_{ii} \over dE}\biggr)\,
 \biggr],
\end{equation}
where {\bf T} is the complex $T$-matrix such that,
$T_{kj} = \Re e T_{kj} \,+ \,i\,\Im m T_{kj}$.
The above relations were put
to a test in \cite{weall} to characterize the hadron resonances
occurring in meson-nucleon and meson-meson elastic scattering. The energy
distribution of the time delay evaluated in these works, nicely displayed
the known $N$ and $\Delta$ excitations and meson resonances 
like the $\rho$, scalars ($f_0$)
and strange $K^*$'s found in $K \pi$ scattering, in addition to confirming
some old claims of exotic states. Evidence for the low-lying 
exotic pentaquark was also found in addition to several other $Z^*$'s from
the time delay in $K^+ N$ scattering.

We shall now demonstrate that the time delay concept which has been so 
useful in characterizing resonances, can also be extended to the search 
of bound as well as unstable bound states. The latter signify states with a
negative binding energy but having a finite lifetime. These are sometimes
called as ``quasibound" states in literature. 
We consider the example of an $S$-matrix for the $n-p$ system
written for the case of a square well potential which has the right
parameters to produce the binding energy of the deuteron. The $S$ matrix
in this case, as a function of $l$ is given as,
\begin{equation}
S_l = - {\alpha h_l^{(2)'}(\alpha) j_l(\beta) - \beta h_l^{(2)}(\alpha)
 j_l{'}(\beta) \over 
\alpha h_l^{(1)'}(\alpha) j_l(\beta) - \beta h_l^{(1)}(\alpha) j_l^{'}(\beta)}
\end{equation}
where $j_l$, $h_l^{(1)}$ and $h_l^{(2)}$ are the spherical Bessel and Hankel
functions of the first and second kind respectively.
$\alpha = k R \,\,\,\,\,{\rm and}\,\,\,\,\, 
\beta = (\alpha^2 - 2 \mu U R^2/\hbar^2)^{1/2}$
where the potential $U$ is given by
$-U = V + i W$ , with $U(r) = U \,\theta (R-r)$
and $R$ the width of the potential well.
A similar square well potential was used by Morimatsu and Yazaki
while locating the ``unstable bound states" of $\Sigma$-hypernuclei as
second quadrant poles and by J. Fraxedas and J. Sesma using the time
delay method \cite{frax}.

We evaluate the time delay in $n-p$ scattering at negative energies
using the above $S$-matrix with $\alpha = i k R$ (hence $E = -k^2/2\mu$)
and $l=0$.
If we plug in the appropriate parameters for an $n-p$
square well potential, $V = 34.6$ MeV and $R = 2.07$ fm, the time delay
plot (as in Fig. 1) shows a sharp spike exactly at the
binding energy of the deuteron ($E=-2.224$ MeV). If we add a small
imaginary part to the potential, then this state develops into a 
Breit-Wigner kind of distribution centered around the binding energy
of the deuteron (a fictitious ``unstable bound state" of the $n-p$
system around $E = -2.224$ MeV).
\begin{figure}
\begin{center}
\includegraphics[width=6cm,height=6cm]{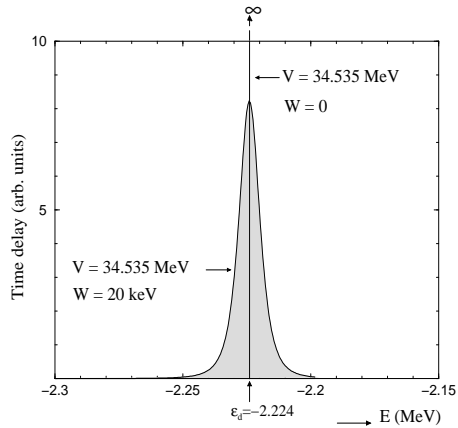}
\caption{\label{fig:epsart} The theoretical delay time in $n p$ 
elastic scattering as a function of the energy $E = \sqrt s - m_n - m_p$, 
where $\sqrt s$ is the total energy available in the $n p$ centre of 
mass system. The sharp spike at $E = -2.224$ MeV corresponding to the 
deuteron binding energy, indicates the infinite time delay due to the 
formation
of this $n p$ bound state. The shaded distribution corresponds to the 
time delay in the case of a fictitious unstable bound state.}
\end{center}
\end{figure}
We can thus see that time delay is well-defined at
negative energies and does reveal the bound and ``unstable bound states"
with positive peaks at negative energies. 
At this point, we may add that in contrast to the infinite 
time delay that occurs in the 
case of bound states, the delay time for virtual states (also called
``antibound" states) is $- \infty$. 
Below, we describe our approach to construct the complex $T$-matrix 
for $\eta \,^3$He elastic scattering which is required for the 
calculation of time delay.

The $\eta \,^3$He transition matrix is evaluated using few body equations 
solved within a Finite Rank Approximation (FRA) approach, which means that the 
$^3$He nucleus in $\eta \,^3$He elastic
scattering remains in its ground state, in the intermediate state. 
Since the $\eta$-mesic bound states
and resonances are basically low energy phenomena, it seems justified to use
the FRA for calculations of the present work. 
Since at low energies, the 
$\eta N$ interaction is dominated by the $S_{11}$ resonance, we restrict
ourselves to the study of the s-wave unstable states. The $\eta \,^3$He 
$t$-matrix in this approach is written as \cite{bela,bela2},
\begin{eqnarray}\label{tfsi}
t_{\eta A}(\vec{k^\prime},\, \vec{k}\,; z) &=& <\, \vec {k^\prime}\, ; \,
\psi_0\,|\, t^0(z)
\, | \, \vec{k} \, ; \, \psi_0\,> \, +\, \\ \nonumber
&&\varepsilon\, \int {\vec{dk^{\prime\prime}}
\over (2\pi)^3} {<\,\vec{k^\prime}\, ; \,  \psi_0 \,|\, t^0(z)\, | \,
\vec{k^{\prime\prime}}\,
; \,  \psi_0\,> \over (z - {k^{\prime\prime\,2} \over 2\mu})(z -
\varepsilon
- {k^{\prime\prime\,2} \over 2\mu})}
t_{\eta A}(\vec{k^{\prime\prime}},\, \vec{k}\, ; \, z) \, ,
\end{eqnarray}
where $z = E - |\varepsilon| + i0$. $E$ is the energy associated with
$\eta A$ relative motion, $\varepsilon$ is the binding energy of 
the nucleus, $\psi_0$ is the nuclear wave function 
and $\mu$ is the reduced mass of the $\eta A$ system. 
The matrix elements for $t^0$ are given as,
\begin{equation}\label{t0mat}
<\, \vec{k^\prime} \, ; \,  \psi_0\,|\, t^0(z)\, |\, \vec {k} \, ; \,
\psi_0\,> =
\int d\vec{r}\, |\, \psi_0(\vec{r})\, |^2 \, t^0\, (\vec{k^\prime},\,
\vec{k}\,;
\vec{r}\,;z)
\end{equation}
where,
\begin{equation}\label{t0mat2}
t^0\,(\vec{k^\prime},\, \vec{k} \,;\vec{r} \,;z) = \sum_{i=1}^A \,
t_i^0\,
(\vec{k^\prime},\, \vec{k}\,;\vec{r_i}\,;z) \, .
\end{equation}
$t_i^0$ is the t-matrix for the scattering of the $\eta$-meson from the
$i^{th}$ nucleon in the nucleus, with the rescattering
from the other (A-1) nucleons included. It is given as,
\begin{equation}\label{t0mat3}
t_i^0\,(\vec{k^\prime},\, \vec{k}\,;\vec{r_i}\,;z) =
t_i^{\eta N}(\vec{k^\prime},\, \vec{k}\,;\vec{r_i}\,;z) + \int
{d\vec{k^{\prime\prime}}
\over (2\pi)^3}\,{t_i^{\eta N}(\vec{k^\prime},\,
\vec{k^{\prime\prime}}\,;\vec{r_i}\,;z)
\over z - {k^{\prime\prime\,2} \over 2\mu}} \sum_{j\neq i}
t_j^0(\vec{k^{\prime\prime}},\, \vec{k}\,;\vec{r_j}\,;z) \, .
\end{equation}
The t-matrix for elementary $\eta$-nucleon scattering, $t_i^{\eta N}$,
is written in terms of the two body $\eta N$ matrix
$t_{\eta N \rightarrow \eta N}$ as,
\begin{equation}\label{tetan}
t_i^{\eta N}(\vec{k^\prime},\, \vec{k}\,;\vec{r_i}\,;z) =
t_{\eta\,N \rightarrow \eta N}(\vec{k^\prime},\,
\vec{k}\,;z)\, exp [\,i (\, \vec{k} -
\vec{k^\prime}\,)\cdot\,\vec{r_i}\,] \, .
\end{equation}
Since there exists a lot of uncertainty
in the knowledge of the $\eta$-nucleon interaction, 
we use three different prescriptions \cite{bhal,fix,green} of
the coupled channel $\eta$-N t-matrix, 
t$_{\eta \, N \, \rightarrow \, \eta \, N}$, leading to different values of 
the $\eta N$ scattering length.

Before proceeding further, we briefly list the general features of the 
present approach, the approximations made and their validity in the
context of the present calculations:\\
(i) The main idea of the FRA consists of separating the motion of
the projectile and the internal motion of the nucleons inside the nucleus.
The total Hamiltonian is split accordingly and the approximation in solving
these equations as mentioned above is to restrict the spectral 
decomposition of the nuclear Hamiltonian to the ground state. 
This, however, limits the applicability of the present approach to energies
below the break up threshold of the $^3$He nucleus. \\
(ii) The operator $t^0$ describes the
scattering of the $\eta$ meson from nucleons fixed in their space position
within the nucleus. However, it differs from the usual fixed center 
t-matrices as $t^0$ is taken off the energy shell and involves the motion of
the $\eta$ meson with respect to the center of mass of the target. 
The present scheme should not be
confused with an optical potential approach which involves
an impulse approximation and the omission of higher order rescattering
terms.
Besides, the $t$-matrix is evaluated at $z = E - |\varepsilon| + i0$ 
in contrast to the usual fixed scatterer approximations which do not
involve the binding energy term. \\
(iii) the dominance of the $N^*-S_{11}$ resonance 
which decays with a similar branching ratio to both the $\pi N$ and
$\eta N$ channels implies that the 
transitions $\eta N \rightarrow \pi N$ must be taken into
account. In the present work we use coupled channel $t$-matrices for 
the elementary $\eta$-nucleon interaction which include fully the
effect of the $\pi N$ and $\eta N$ channel on each other. 
The complex self energy terms in the $\eta N \to \eta N$ t-matrix take 
care of the intermediate off shell as well as on shell $\pi N$ loops. 
We neglect however, transitions involving the production of an 
$\eta$ meson on a second nucleon by a pion produced on the first nucleon. 
For example, in the present three nucleon case, we neglect processes
of the type, $\eta N_1 \to \pi N_1$ followed by $\pi N_2 \to \eta N_2$ or
$\pi N_2 \to \pi N_2$ followed by $\pi N_2 \to \eta N_2$ etc.  
These are in principle 
more difficult to calculate. Though this neglect cannot be justified
a priori, there is reason enough to believe that these 
contributions may not be significantly large as argued in \cite{fix3}.\\
(iv) The $^3$He nuclear wave function required in the
calculation of the $T$-matrix is taken to be of the Gaussian form. 
The use of this wave function for the present work seems
adequate in the light of the 
calculations performed in \cite{bela2} where 
searches for eta-mesic states using a more sophisticated
wave function for $^3$He resulted in almost the same locations 
of the $T$-matrix poles as found using the Gaussian one.

The $T$-matrix for $\eta A$ elastic scattering, $t_{\eta A}$, is related to
the $S$-matrix as,
\begin{equation}\label{smat}
S\,=\, 1 \,-\, {\mu \,i\,k \over \pi}\,t_{\eta A} \, ,
\end{equation}
where $k$ is the momentum in the $\eta A$ centre of mass system
and hence, the dimensionless $T$-matrix required in the evaluation of
time delay is given as,
$T=-\,(\mu \,k / \,2\,\pi) \,t_{\eta A}$.
and used to evaluate the time delay given in 
(\ref {7}) for the reaction $\eta \,^3$He $\rightarrow \eta \,^3$He. 
We evaluate $\Delta t_{ii} (E)$ at both positive and negative values
of $E$ in order to search for resonances and unstable bound states.
As explained above with the example of the deuteron, time delay at
negative energies is a valid concept and is useful in identifying the
bound and unstable bound states. However, 
in contrast to positive energy resonances where one can 
``measure" the time delay from phase shifts in a scattering experiment, 
the time delay at negative energies 
is not measurable in an elastic scattering experiment.  
At negative energies however, it manifests in an off-shell elastic scattering,  
where the transition matrix
$t_{\eta A}$ is calculated at purely imaginary momenta 
($ k = k^{\prime} \to i k$) corresponding to $E = -k^2/2\mu$ with $\mu$ being
the reduced mass of the $\eta \,A$ system. Physically, this would
correspond to the binding of subthreshold produced $\eta$'s in 
nuclear reactions through off shell scattering.

\begin{figure}[ht]
\begin{center}
\includegraphics[width=10cm,height=8cm]{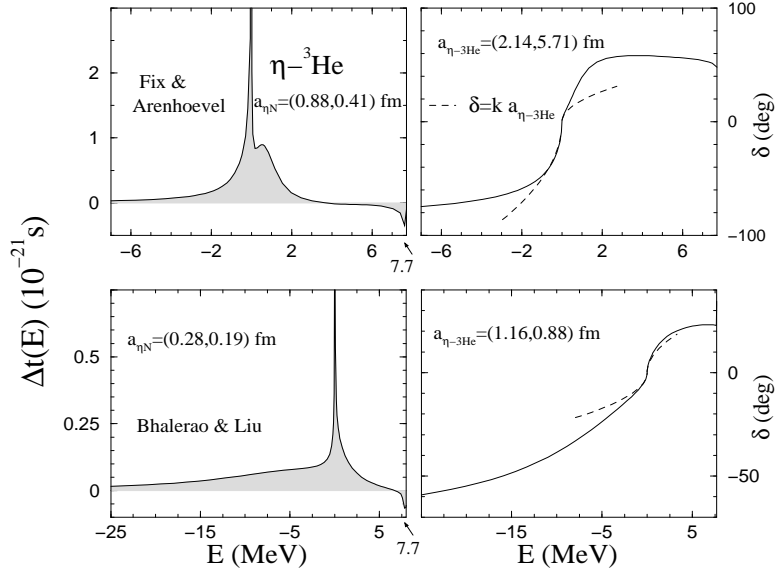}
\caption{\label{fig:eps2} Time delay in the elastic scattering reaction, 
$\eta \,^3$He $\rightarrow \eta \,^3$He, as a function of energy, 
$E = E_{\eta \,^3He} - m_{\eta} - m_{^3He}$, where $E_{\eta \,^3He}$ is the
total energy available in the $\eta \,^3$He centre of mass. 
The vertical axis scale has been broken to display the
structure of the time delay plot clearly.}
\end{center}
\end{figure}
In Fig. 2, we show the $\eta \,^3$He time delay plots 
with two different inputs for the $\eta N$ interaction \cite{bhal,fix}. 
In \cite{fix} a coupled channel t-matrix including the $\pi$N and
$\eta$N channels with the S$_{11}$ - $\eta$N interaction playing a
dominant role was constructed. It consisted of the
meson - N* vertices and the N* propagator as given below:
\begin{equation}\label{tfix}
t_{\eta \, N \, \rightarrow \, \eta \, N} (\, k^\prime, \, k; z) = 
{ { \rm g}_{_{N^*}}\beta^2 \over (k^{\prime\,2} +
\beta^2)}\,\tau_{_{N^*}}(z)\,{ {\rm g}_{_{N^*}}\beta^2 \over (k^2 + \beta^2)}
\end{equation}
with,
$\tau_{_{N^*}}(z) = ( \, z - M_0- \Sigma_\pi(z) - 
\Sigma_\eta(z) + i\epsilon \, )^{-1}$, 
where $\Sigma_\alpha(z)$ $(\alpha = \pi, \eta)$ are the self energy
contributions from the $\pi N$ and  $\eta N$ loops. 
The parameters were fitted such that they
reproduced scattering lengths of $a_{\eta N}$ = (0.75, 0.27) fm and 
$a_{\eta N}$ = (0.88, 0.41) fm, which are in 
agreement with values obtained from modern analyses of recent data. 
We chose the parameter set with  
${\rm g}_{_{N^*}}=2.13, \beta=13 \,{\rm fm}^{-1}$ and 
$M_0 = 1656 \,{\rm MeV} $, which gave rise to 
$a_{\eta N}$ = (0.88, 0.41) fm. 
In \cite{bhal}, the $\pi$N, $\eta$N and 
$\pi \Delta$ ($\pi \pi N$) channels were treated
in a coupled channel formalism (so that an additional self-energy
term appears in the propagator $\tau_{_{N^*}}(z)$).
The parameters of this model are,
${ \rm g}_{_{N^*}}=0.616$, $\beta=2.36 \,{\rm fm}^{-1}$ and 
$M_0 = 1608.1 \,{\rm MeV}$. These parameters were obtained from a fit
to the $P_{33}$ and $S_{11}$ phase shifts in $\pi N$ elastic scattering 
and the differential $\pi N \rightarrow \eta N$ 
cross sections predicted by the model were in good agreement with data.
The input parameters of this model, however, give rise to a much
smaller scattering length, namely, $a_{\eta N}$ = (0.28, 0.19) fm. 
It should be noted that though the two models predict somewhat 
different total $\pi N \rightarrow \eta N$ cross sections, they 
both agree with data since the error bars on the data are large.

In both cases we find a peak centered slightly below
zero energy. In the case of the larger $\eta N$ scattering length \cite{fix}, 
we find an additional bump 
located at a positive energy of $0.5$ MeV and overlapping the peak 
very close to threshold. In the case of the smaller $a_{\eta N}$ 
(lower half), there is a hint of a broad hump around $-5$ MeV. 
These findings seem to be in agreement 
with experiment where a binding energy of $-4.4 \pm 4.2$ MeV for 
the $\eta$-mesic $^3$He and a resonance like structure just above 
production threshold was reported.

Before proceeding further, we would like to mention that 
in the case of s-wave scattering near the elastic threshold, 
i.e. at zero energy, one needs to be careful in drawing conclusions
from time delay plots. Using $S = exp(2i\delta)$ 
and comparing it with (\ref{smat}),
\begin{equation}
\delta = {1 \over 2i} \, {\rm ln}(1 - {i \mu k \over \pi} t_{\eta A}) \, = \,
{1 \over 2i} \, {\rm ln}(1 + 2 i k f) 
\end{equation}
where $f$ is the scattering amplitude. For small $k$, $\delta \simeq k f$ and
the behaviour of $d\delta/dE$ (the real part of 
which is essentially the time delay) is
determined by the simple pole at $k=0$ (or $E_{\eta ^3He} = E_{threshold}$) and 
the energy dependence of the scattering amplitude $f$. In the absence of
a resonance, as $k \to 0$, $\delta = k a$, where $a = a_R + i a_I$ 
is the complex scattering length. For positive energies, 
$\Re e \delta = k a_R$, whereas for energies below zero, $ k \to ik$ and
$\Re e \delta = - k a_I$. In such a situation, $\Re e (d\delta/dE)$ exhibits 
a sharp peak at $E_{\eta^3He} = E_{threshold}$, the sign of which is determined by the 
sign of the scattering length. 
On the other hand, if the scattering amplitude has a resonant behaviour
near threshold, one would see a superposition of the two behaviours. 
For a state far from threshold, the rise
in time delay near threshold is extremely sharp and the shape of
the state remains completely unaffected. A state reasonably close
to threshold is distorted in shape and one very close 
manifests simply by broadening the threshold singularity. 
In Fig. 2, we display the phase shifts and 
their threshold behaviour (dashed lines) 
which depend on the $\eta-^3$He scattering lengths. These phase shifts
do display the characteristic resonant behaviour apart from their 
agreement with $\delta = k a$ close to threshold.

The formalism of the present work forces the three nucleon
system to be in the bound state always and hence searching for
unstable states beyond the $^3$He break up threshold of 7.7 MeV as
well as below -7.7 MeV is beyond the reach of the present approach. 
This limitation taken along with the fact that virtual target
excitations could be important below the break up threshold add 
some uncertainty to the bump around $-5$ MeV. However, this 
bump is quite broad and spreads over to the positive 
energy side near threshold where the FRA is reliable. 
Besides, we note that the $\eta N$ scattering length in this case
is very small ($a_{\eta N} = (0.28, 0.19)$ fm) and in a comparison 
of the Alt-Grassberger-Sandhas (AGS) equations
(which include the target excitations) with FRA, the authors in 
\cite{shevd} found that 
the FRA works reasonably well for real parts of $a_{\eta N} < 0.5$ fm. 
The above peak is also expected to be distorted by the singularity in the
threshold region as mentioned above. Thus, one cannot be sure about 
the exact peak position of this state, but the plot does seem to 
indicate the existence of one broad state in this region within
the eta-nucleon model of Bhalerao and Liu.

Another interesting feature in these time delay plots is the 
occurrence of negative time delay around $7.7$ MeV which is essentially
the binding energy of the $^3$He nucleus and is also an input to
our calculation (see Eq. \ref{tfsi}). It can be understood by noting 
the connection of the density of
states with the energy derivative of the phase shift which also defines 
time delay. According to the Beth-Uhlenbeck formula \cite{beth}, 
the difference in the density of states with and without interaction is
given by the energy derivative of the phase shift. This means that when
the density of states due to interaction is less than that without
interaction, the energy derivative of the phase shift would be negative. 
The reduction in the density of states due to interaction
in our context corresponds to a loss of flux from the elastic channel
due to the opening of an inelastic channel (more examples in \cite{me}).

\begin{figure}[ht]
\begin{center}
\includegraphics[width=10cm,height=8cm]{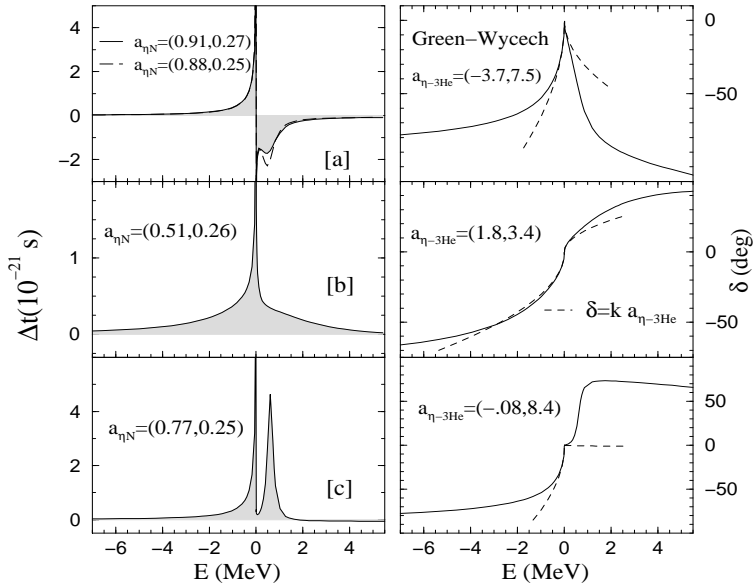}
\caption{\label{fig:eps3} Same as Fig. 2, but with a different 
model of the $\eta N$ interaction. Scattering lengths are given in fm.}
\end{center}
\end{figure}
In Fig. 3, we show the time delay plots 
and corresponding phase shifts using a simple separable form of
the off-shell $t$-matrix \cite{green}. Though this $t$-matrix 
does not contain the self-energy contributions as in \cite{bhal,fix}, 
it has the advantage that it is parametrized
with the latest Crystal Ball data on $\pi N \rightarrow \eta N$ \cite{crystal} 
and also includes the $\gamma N \rightarrow \eta N$ 
data \cite{gam}. In Fig. 3a, we observe subthreshold peaks for two
sets of $t$-matrix solutions (Set A and B from \cite{green}), 
out of which the solid line
indicates their best solution and dashed one a similar solution without
pion beam momentum correction as explained in \cite{green}. 
In both these cases, however, we observe negative time delay above
threshold which could be due to a repulsive interaction \cite{smith,me}. 
In Fig. 3b, one sees a broad peak at threshold, very similar to
those in Fig. 2. This result corresponds to Set D in \cite{green} which
is an unconventional solution obtained when the recent data 
from \cite{crystal} is omitted. Finally, Fig. 3c displays two distinct
peaks using the Set E solution which omits the data from 
\cite{gam}.        
Though we have listed only the scattering lengths corresponding to
the various solutions, it should be noted that the results are also
sensitive to the effective range parameter $r_0$ given in Table III of
\cite{green}.

Before concluding, we mention an analysis \cite{sibir} where, 
using optical potentials, the authors investigated the relation 
between the $\eta$-$^3$He binding
energy, width and the complex scattering length, $a_{\eta-^3He}$. 
The strength of the potential was varied to check for which 
scattering lengths one can find bound states. 
They found that if an $\eta$-mesic $^3$He state exists,
its binding energy (width) should not exceed $5$ MeV ($10$ MeV). 
It is interesting to note that though the approach of the present
work which involves the indentification of the quasi bound state 
from positive peaks in time delay is entirely different from that
in \cite{sibir}, 
the possible $\eta$-mesic states of the present work lie within
this limit.  
The constraints on the energy and width in \cite{sibir} 
were found following 
a systematic analysis of the $\eta$-$^3$He scattering 
length \cite{sibir2}. 
The values of $a_{\eta-^3He}$ obtained in the present work 
are outputs of the few body calculations for a given 
$\eta N$ interaction that enters the $\eta$-nucleus $t$-matrix 
as an input. 
They are listed in Figs 2 and 3. We refer the reader to \cite{sibir2} and
the references therein for a detailed discussion on the 
issue of the $\eta$-nucleus scattering lengths.

In conclusion, we have investigated the possible existence of 
$\eta$-mesic $^3$He using the physical concept of time delay 
and within the limitations of the present approach, 
find evidence for the existence of such states near threshold
using two off-shell isobar models (\cite{bhal} and \cite{fix}) of the 
$\eta N$ interaction and another $K$-matrix semiphenomenological analysis
of relevant data \cite{green}. 
With the availability of more accurate data on $\eta$-mesic nuclei, 
the present results will be useful in obtaining a better
knowledge of the $\eta N$ interaction. 
 

\end{document}